\documentclass{article}
\usepackage{amsmath}
\usepackage{cite}
\usepackage{graphicx}
\usepackage{dcolumn}

\begin{document}

\date{}
\title{On the exact solutions of a one-dimensional Schr\"odinger equation with a
rational potential}
\author{Francisco M. Fern\'{a}ndez\thanks{%
fernande@quimica.unlp.edu.ar} \\
%EndAName
INIFTA, DQT, Sucursal 4, C. C. 16, \\
1900 La Plata, Argentina}
\maketitle

\begin{abstract}
We analyse the exact solutions of a conditionally-solvable Schr\"odinger
equation with a rational potential. From the nodes of the exact
eigenfunctions we derive a connection between the otherwise isolated exact
eigenvalues and the actual eigenvalues of the Hamiltonian operator.
\end{abstract}

\section{Introduction}

\label{sec:intro}

Some time ago, Mitra\cite{M78} discussed a one-dimensional Schr\"{o}dinger
equation with a rational potential that exhibits interest in some fields of
physics. Somewhat later, Flessas\cite{F81} derived exact solutions for some
particular values of the model parameters. This fact made the problem quite
popular and other authors discussed the exact solutions with somewhat more
detail\cite
{V81,WWN82,LL82,H83,CM83,BL87,RR87,RRV88,G88,L89,VD89,BV89,RR90,ADV91,PM91,SHC06}%
. It was shown that one can obtain exact solutions for that equation by
means of a simple power-series method\cite
{F81,V81,WWN82,LL82,H83,CM83,BL87,G88,L89,VD89,BV89,RR90}, an approach that
is facilitated by the use of tridiagonal matrices and their determinants\cite
{WWN82,H83,CM83,VD89,BV89}. Such exact solutions can also be derived by
means of supersymmetric quantum mechanics\cite{RR87,RRV88,RR90,ADV91} and
the dynamical symmetry of the problem was also investigated\cite{RR90}. It
is also possible to carry out the analysis by means of the Heun confluent
equation\cite{PM91}. Other authors also considered two- three- and general $N
$-dimensional models\cite{BV89,PM91}. Particular exact solutions for a
non-exactly solvable quantum-mechanical model are suitable for testing
approximate methods\cite{SHC06}.

The rational potentials mentioned above are particular cases of
conditionally-solvable problems\cite{CDW00,T16} (and references therein) and
the approximate solutions of some of them have been misinterpreted by many
authors in supposedly physical applications on a wide variety of fields\cite
{F21} (and references therein).

The purpose of this paper is the analysis of some features of the
distribution of the exact eigenvalues of the Schr\"{o}dinger equation with
the one-dimensional rational potential that passed unnoticed in the papers
mentioned above.

In section~\ref{sec:model} we present the model and derive a suitable
dimensionless equation. In section~\ref{sec:TTRR} we discuss the exact
solutions produced by the power-series method and organize the eigenvalues
in a suitable way for their interpretation. Finally, in section~\ref
{sec:conclusions} we summarize the main results and draw conclusions.

\section{The model}

\label{sec:model}

The model discussed here is given by the Hamiltonian operator
\begin{equation}
H=-\frac{\hbar ^{2}}{2m}\frac{d^{2}}{dx^{2}}+V_{1}x^{2}+\frac{V_{2}x^{2}}{%
x^{2}+x_{0}^{2}},  \label{eq:H}
\end{equation}
where $-\infty <x<\infty $, $V_{1}>0$ and $-\infty <V_{2}<\infty $. In order
to obtain a dimensionless eigenvalue equation we choose a suitable unit of
length $L$ and define the dimensionless coordinate $\tilde{x}=x/L$. The unit
of length $L$ is arbitrary but it is commonly convenient to use it for
removing the physical constants and some chosen model parameters\cite{F20}.
For example, if we want to get rid of $V_{1}$ we choose $L=\left[ \hbar
^{2}/\left( 2mV_{1}\right) \right] ^{1/4}$ so that the resulting
dimensionless Hamiltonian operator becomes
\begin{equation}
\tilde{H}=\frac{2mL^{2}}{\hbar ^{2}}H=-\frac{d^{2}}{d\tilde{x}^{2}}+\tilde{x}%
^{2}+\frac{\lambda \tilde{x}^{2}}{1+g\tilde{x}^{2}},\;\lambda =\frac{V_{2}}{%
V_{1}x_{0}^{2}},\;g=\frac{\hbar }{\sqrt{2mV_{1}}x_{0}^{2}},  \label{eq:H_dim}
\end{equation}
and the unit of energy is $\hbar ^{2}/(2mL^{2})=\hbar \sqrt{V_{1}/(2m)}$.
From now on we remove the tilde on the dimensionless operators.

\section{Exact polynomial solutions}

\label{sec:TTRR}

In order to obtain polynomial solutions to the dimensionless Schr\"{o}dinger
equation $H\psi =E\psi $ we resort to the ansatz
\begin{equation}
\psi (x)=x^{s}e^{-x^{2}/2}\sum_{j=0}^{\infty }c_{j}x^{2j},  \label{eq:psi}
\end{equation}
where $s=0$ for even states and $s=1$ for odd ones. The expansion
coefficients $c_{j}$ satisfy the three-term recurrence relation
\begin{eqnarray}
c_{j+2} &=&A_{j}c_{j+1}+B_{j}c_{j},\;j=0,1,\ldots ,  \nonumber \\
A_{j} &=&-\frac{E+2g\left( j+1\right) \left( 2j+2s+1\right) -4j-2s-5}{%
2\left( j+2\right) \left( 2j+2s+3\right) },  \nonumber \\
B_{j} &=&-\frac{Eg-g\left( 4j+2s+1\right) -\lambda }{2\left( j+2\right)
\left( 2j+2s+3\right) }.  \label{eq:TTRR}
\end{eqnarray}
If $c_{n}\neq 0$ and $c_{n+1}=c_{n+2}=0$, $n=0,1,\ldots $, then $c_{j}=0$
for all $j>n$ and the power-series expansion in the ansatz (\ref{eq:psi})
reduces to a polynomial of degree $n$. These conditions require that $B_{n}=0
$ from which we obtain
\begin{equation}
E=4n+2s+1+\frac{\lambda }{g}.  \label{eq:E}
\end{equation}
This expression cannot be interpreted as the eigenvalues of the
quantum-mechanical model (except in the trivial case $\lambda =0$) as
discussed in what follows. The coefficients $A_{j}$ and $B_{j}$ of the
recurrence relation (\ref{eq:TTRR}) become
\begin{eqnarray}
A_{j} &=&-\frac{2g^{2}\left( j+1\right) \left( 2j+2s+1\right) -4g\left(
j-n+1\right) +\lambda }{2g\left( j+2\right) \left( 2j+2s+3\right) },
\nonumber \\
B_{j} &=&\frac{2g\left( j-n\right) }{\left( j+2\right) \left( 2j+2s+3\right)
}.  \label{eq:Ajn,Bjn}
\end{eqnarray}

Equation (\ref{eq:E}) does not give us the energy eigenvalues for all
allowed values of the model parameters because the condition $%
c_{n+1}(\lambda ,g,s)=0$ makes the exact polynomial solutions to be valid
only for particular values of $\lambda $ and $g$. For example, when $n=0$ we
have
\begin{equation}
c_{1}=-\frac{\lambda }{2g\left( 2s+1\right) }=0,  \label{eq:c_1}
\end{equation}
that leads to $\lambda =0$. In this case we only obtain a solution for the
harmonic oscillator with energy $E=2s+1$. The usual notation for the
eigenvalues of a one-dimensional quantum-mechanical model is $E_{\nu }$,
where $\nu =0,1,\ldots $ is the number of nodes of the corresponding
wavefunction $\psi _{\nu }(x)$. In this case it is clear that $\nu =s$,
where $s=0$ (no nodes) and $s=1$ (one node at $x=0$ because of the factor $%
x^{s}$ in equation (\ref{eq:psi})).

When $n=1$ the condition
\begin{equation}
c_{2}=\frac{\lambda \left( 2g^{2}\left( 2s+1\right) +4g+\lambda \right) }{%
8g^{2}\left( 2s+1\right) \left( 2s+3\right) }=0,  \label{eq:c_2}
\end{equation}
exhibits two solutions
\begin{equation}
\lambda ^{(1,1)}=-2g\left[ g\left( 2s+1\right) +2\right] ,\;\lambda
^{(1,2)}=0,  \label{eq:lambda^(1,i)}
\end{equation}
with polynomials
\begin{equation}
p^{(1,1)}(x)=1+gx^{2},\;p^{(1,2)}(x)=1-\frac{2x^{2}}{2s+1}.
\label{eq:p^(1,i)}
\end{equation}
Note that $p^{(1.1)}(x)$ is nodeless and $\nu =s$; on the other hand, $%
p^{(1,2)}(x)$ exhibits two nodes and $\nu =2+s$. At this point it is clear
that the number $n$ is the degree of the polynomial factor but it is not
related to the number of zeros of the wavefunction.

When $n=2$ the condition
\begin{equation}
c_{3}=-\frac{\lambda \left( 8g^{4}\left( 2s+1\right) \left( 2s+3\right)
+32g^{3}\left( 2s+3\right) +2g^{2}\left( \lambda \left( 6s+7\right)
+16\right) +12g\lambda +\lambda ^{2}\right) }{48g^{3}\left( 2s+1\right)
\left( 2s+3\right) \left( 2s+5\right) }=0,  \label{eq:c_3}
\end{equation}
exhibits three roots
\begin{eqnarray}
\lambda ^{(2,1)} &=&-g\left[ \sqrt{g^{2}\left( 4s^{2}+20s+25\right)
+4g\left( 2s-3\right) +4}+g\left( 6s+7\right) +6\right] ,  \nonumber \\
\lambda ^{(2,2)} &=&g\left[ \sqrt{g^{2}\left( 4s^{2}+20s+25\right) +4g\left(
2s-3\right) +4}-g\left( 6s+7\right) -6\right] ,  \nonumber \\
\lambda ^{(2,3)} &=&0,  \label{eq:lambda^(2,i)}
\end{eqnarray}
with polynomials
\begin{eqnarray}
p^{(2,1)}(x) &=&1+\frac{\sqrt{g^{2}\left( 4s^{2}+20s+25\right) +4g\left(
2s-3\right) +4}+g\left( 6s+7\right) -2}{2\left( 2s+1\right) }x^{2}  \nonumber
\\
&&+\frac{\sqrt{g^{2}\left( 4s^{2}+20s+25\right) +4g\left( 2s-3\right) +4}%
+g\left( 2s+5\right) -2}{2\left( 2s+1\right) }gx^{4},  \nonumber \\
p^{(2,2)}(x) &=&1-\frac{\sqrt{g^{2}\left( 4s^{2}+20s+25\right) +4g\left(
2s-3\right) +4}-g\left( 6s+7\right) +2}{2\left( 2s+1\right) }x^{2}  \nonumber
\\
&&-\frac{\sqrt{g^{2}\left( 4s^{2}+20s+25\right) +4g\left( 2s-3\right) +4}%
-g\left( 2s+5\right) +2}{2\left( 2s+1\right) }gx^{4},  \nonumber \\
p^{(2,3)}(x) &=&1-\frac{4x^{2}}{2s+1}+\frac{4x^{4}}{\left( 2s+1\right)
\left( 2s+3\right) },  \label{eq:p^(2,i)}
\end{eqnarray}
with no nodes, two nodes and four nodes, respectively.

In general, the condition $c_{n+1}=0$ yields $n+1$ roots $\lambda
_{s}^{(n,1)}<\lambda _{s}^{(n,2)}<\ldots <\lambda _{s}^{(n,n+1)}=0$ and the
number of nodes of the corresponding wavefunctions is given by $\nu
=2(i-1)+s $. Note that we can obtain exact polynomial solutions only for $%
\lambda \leq 0$ as already reported in all the earlier papers already
mentioned above. The exact eigenvalues can be written as
\begin{equation}
E_{s}^{(n,i)}(g)=4n+2s+1+\frac{\lambda _{s}^{(n,i)}}{g}.
\label{eq:E^(n,i)_s}
\end{equation}
From the analysis just given we arrive at one of the main results of this
paper:
\begin{equation}
E_{s}^{(n,i)}(g)=E_{2(i-1)+s}(\lambda ,g),\;n=0,1,\ldots ,\;i=1,2,\ldots
,n+1,  \label{eq:E_nu}
\end{equation}
where $E_{s}^{(n,n+1)}=E_{\nu }(0)=2\nu +1=4n+2s+1$ is the spectrum of the
harmonic oscillator.

In order to test the analytical expressions derived above we carried out a
numerical calculation of the eigenvalues $E_{\nu }(\lambda ,g)$ by means of
the Rayleigh-Ritz (RR) method with the nonorthogonal basis set of functions $%
\varphi _{j,s}(x)=x^{2j+s}\left( 1+gx^{2}\right) e^{-x^{2}/2}$, $%
j=0,1,\ldots $. The details of this well known approach are given in most
textbooks on quantum chemistry\cite{P68,SO96} and elsewhere\cite{F24}; for
this reason we omit them here. Figure~\ref{Fig:Enu0RR} shows some exact
eigenvalues $E_{0}^{(n,i)}(1)$ as well as the RR eigenvalues $E_{\nu
}(\lambda ,1)$, $\nu =0,2,\ldots ,8$. The latter were obtained with a basis
set of $22$ functions ($j=0,1,\ldots ,21$, $s=0$) and are sufficiently
accurate for present purposes. Figure~\ref{Fig:Enu1RR} shows the same data
for the odd functions ($s=1$). We appreciate that in both cases the RR
eigenvalues $E_{\nu }(\lambda ,1)$ pass through the points $E_{s}^{(n,i)}(1)$
thus giving the latter their true meaning as eigenvalues of the Hamiltonian
operator for each given value of $\lambda $. Note that the RR eigenvalues
have the correct slope according to the Hellmann-Feynman theorem (HFT)\cite
{G32,F39}
\begin{equation}
\frac{\partial E}{\partial \lambda }=\left\langle \frac{x^{2}}{1+gx^{2}}%
\right\rangle >0.  \label{eq:HFT_lambda}
\end{equation}

Figure~\ref{Fig:E1G} shows some exact eigenvalues $E_{0}^{(n,1)}(g)$ for $%
g=0.2,\;0.5,\;1$ and some values of $\lambda $. In this case we appreciate
that these eigenvalues also have the correct slope according to the HFT
\begin{equation}
\frac{\partial E}{\partial g}=-\lambda \left\langle \frac{x^{4}}{\left(
1+gx^{2}\right) ^{2}}\right\rangle >0,\;\lambda >0.  \label{eq:HFT_g}
\end{equation}
It is worth noting that $E_{0}^{(n,1)}(g)=E_{0}(\lambda ^{(n,1)},g)$.

\section{Conclusions}

\label{sec:conclusions}

\begin{figure}[tbp]
\begin{center}
\includegraphics[width=9cm]{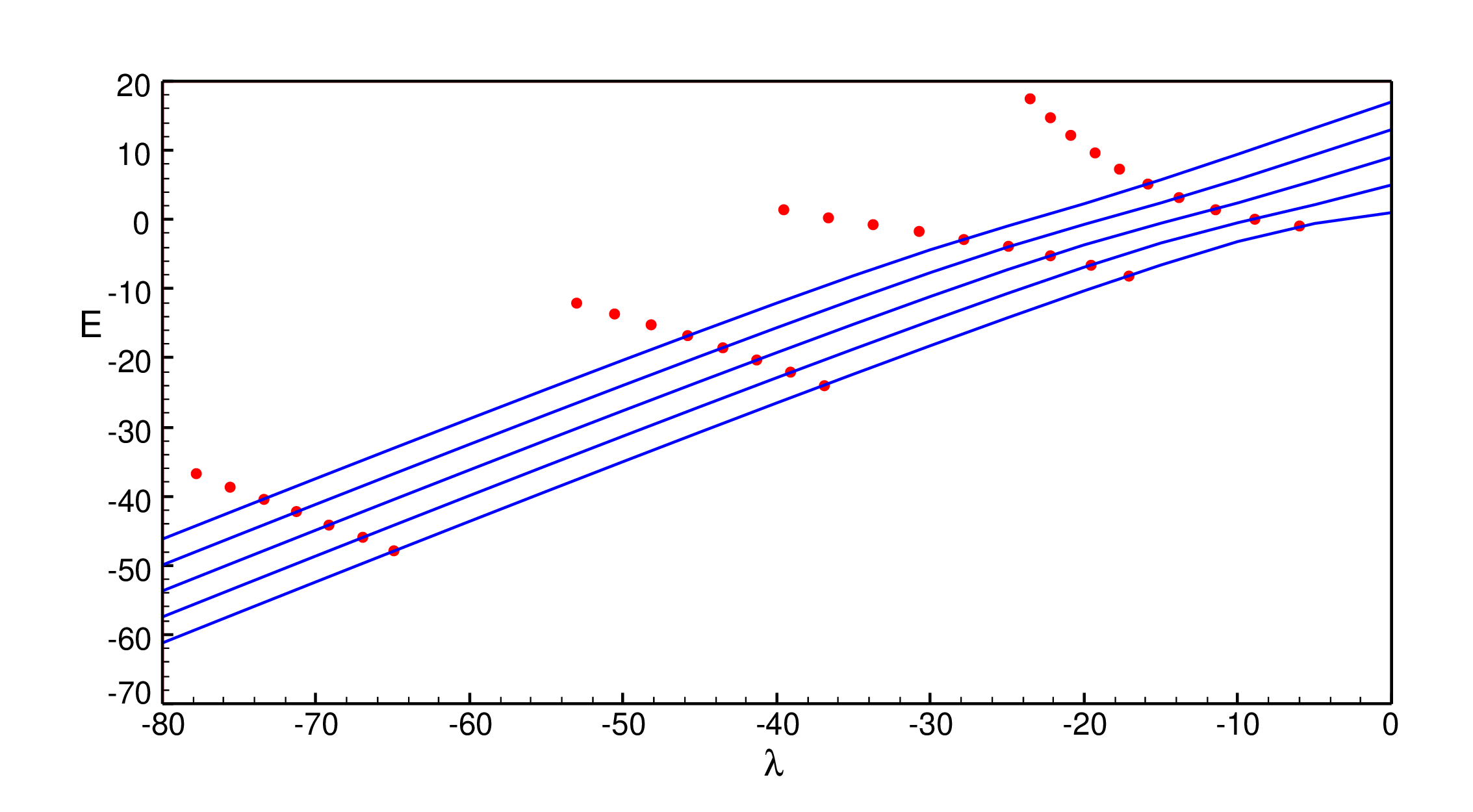}
\end{center}
\caption{Exact eigenvalues $E^{(n,i)}$ (red circles) and RR eigenvalues $%
E_\nu$, $\nu=0,2,\ldots,8$ (blue lines) for $g=1$}
\label{Fig:Enu0RR}
\end{figure}

\begin{figure}[tbp]
\begin{center}
\includegraphics[width=9cm]{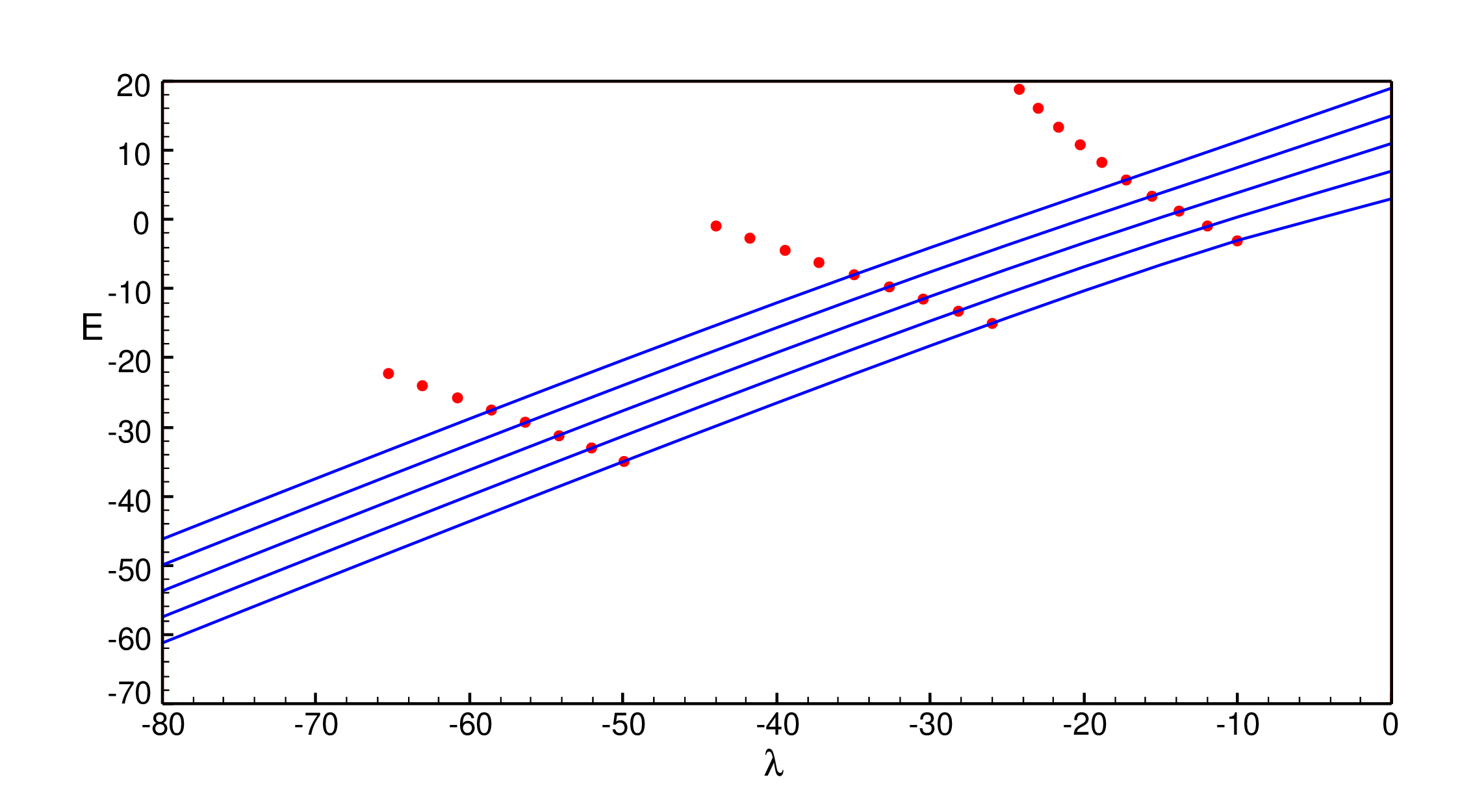}
\end{center}
\caption{Exact eigenvalues $E^{(n,i)}$ (red circles) and RR eigenvalues $%
E_\nu$, $\nu=1,3,\ldots,9$ (blue lines) for $g=1$}
\label{Fig:Enu1RR}
\end{figure}

\begin{figure}[tbp]
\begin{center}
\includegraphics[width=9cm]{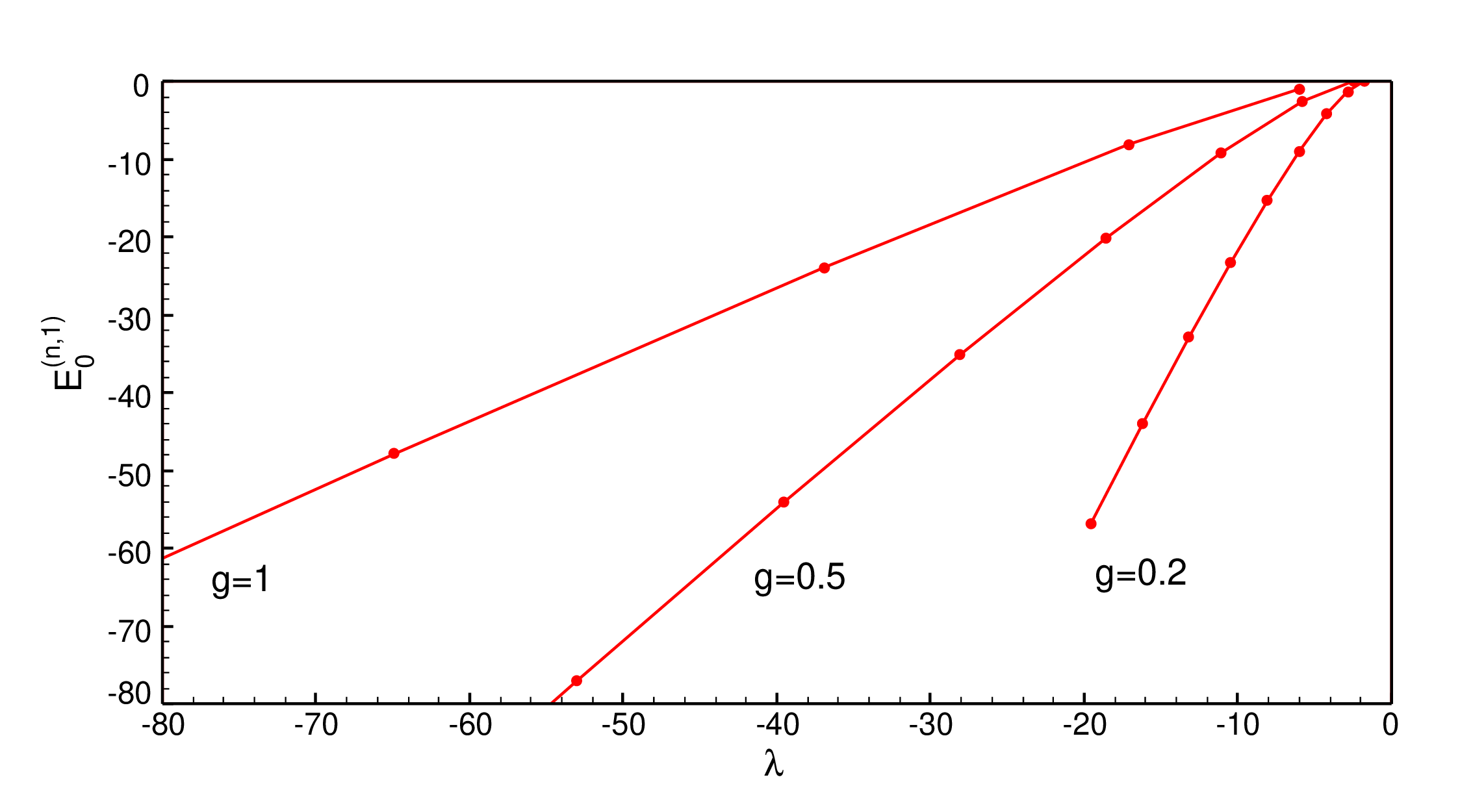}
\end{center}
\caption{Exact eigenvalues $E_0^{(n,1)}(g)$}
\label{Fig:E1G}
\end{figure}

The main purpose of this paper is to show that the exact eigenvalues $%
E_{s}^{(n,i)}$, $n=0,1,\ldots $, $i=1,2,\ldots ,n+1$, associated with the
polynomial solutions are particularly useful and relevant when one can
connect them with the eigenvalues $E_{\nu }(\lambda ,g)$ of the
quantum-mechanical model. The most relevant result is that $\nu =2(i-1)+s$
that enables us to obtain some exact eigenvalues $E_{\nu }(\lambda ,g)$ for
a given value of $g$ and particular values of $\lambda <0$. This fact is
clearly illustrated in figures \ref{Fig:Enu0RR}, \ref{Fig:Enu1RR} and \ref
{Fig:E1G}. If one does not identify the exact eigenvalues $E_{s}^{(n,i)}$%
properly, one is bound to make serious mistakes as discussed elsewhere\cite
{F21}.

\end{document}